\documentclass[aps,prb,twocolumn,superscriptaddress,floatfix]{revtex4}
\usepackage{color,graphicx}
\usepackage{epsfig}
\usepackage{subfigure}
\usepackage{amsmath,amsfonts,amssymb,amsthm,verbatim}
\usepackage{tabularx}
\usepackage{bm}

\def\be{\begin{equation}}
\def\ee{\end{equation}}
\def\ba{\begin{eqnarray}}
\def\ea{\end{eqnarray}}
\def\Cal{\mathcal}

\begin{document}
\title{Novel properties in Josephson junctions involving the $\cos(k_x)\cdot \cos(k_y)$-pairing state in iron-pnictides}
\author{Wei-Feng Tsai}
\affiliation{Department of Physics, Purdue University, West Lafayette, IN 47907, USA}
\author{Dao-Xin Yao}
\affiliation{Department of Physics, Purdue University, West
Lafayette, IN 47907, USA}
\author{B. Andrei Bernevig}
\affiliation{Princeton Center for Theoretical Science, Princeton
University, Priceton, NJ 08544, USA}\author{JiangPing Hu}
\affiliation{Department
of Physics, Purdue University, West Lafayette, IN 47907, USA}
\date{\today}
\newcommand{\br}{\mathbf{r}}
\newcommand{\brprime}{{\mathbf{r}^\prime}}
\newcommand{\bk}{\mathbf{k}}

\begin{abstract}
We propose a novel trilayer $\pi$-junction that takes advantage of the  unconventional $s_{x^2 y^2}=\cos k_x\cos k_y$ pairing symmetry which changes sign between electron and hole Fermi pockets in the iron-pnictides. In addition, we also present theoretical results for Andreev bound states in thin
superconductor-normal metal (or insulator)-iron-pnictide junctions.
The presence of non-trivial in-gap states, which uniquely appear in this unconventional pairing state, is a distinct feature in comparison to other singlet pairing states.
\end{abstract}
\maketitle

\begin{figure}[tbh]
\begin{center}
\includegraphics[scale=0.42]{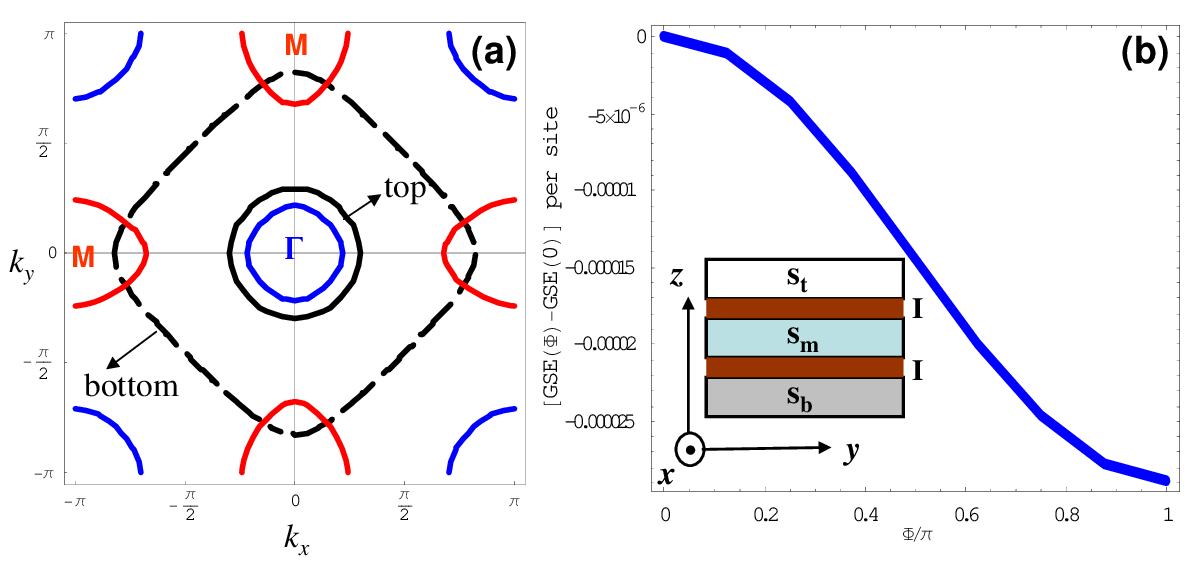}
\caption{(Color online) The plots of (a) schematic stacked Fermi
surfaces from tri-layer superconducting junction and (b) the
ground state energy deviation (from $\Phi=0$ case) per site as a function of relative phase $\Phi$ with $\theta_m =0$. The colored curves in (a) represent the FSs of the iron-pnictides.
%The blue and red curves in (a) represent hole and electron pockets for iron-pnictides respectively.
} \label{fig:piJunc}
\end{center}
\end{figure}

%$Introduction$
A family of iron-based high-temperature superconductors
has recently been discovered.\cite{kamihara08}  These compounds triggered enormous experimental and theoretical interest. In particular, probing the Cooper pair symmetry is critical to understanding the pairing mechanism of this new type of superconductors. Theoretically, many possible gap pairing symmetries have been proposed for iron pnictides, due to the material's multi-orbital nature and complex Fermi surfaces (FSs), with two hole pockets around $\Gamma$-point and two electron pockets around $M$-point [see Fig.~\ref{fig:piJunc} (a)].

Among all the candidates, the proposal of $s$-wave pairing symmetry
with relative sign change between hole and electron pockets has
appealing advantages.\cite{mazin08,seo08,wang08,cvetkovic08} Two of us have predicted,\cite{seo08} based on a local magnetic exchange coupling $J_1$-$J_2$
model,\cite{yildirim,Ma,si,Fang2008} an unconventional $s$-wave symmetry with a
particular $s_{x^2y^2}=\cos k_x\cos k_y$ form in the reciprocal
momentum space. The predicted order parameter is consistent with the relative values of the gap on the hole and electron FSs reported by angle-resolved photo-emission spectroscopy
(ARPES) experiments.\cite{Ding} The proposed symmetry is
also consistent with low temperature-dependent penetration depth
experiments,\cite{martin08,hashimoto} and partially explains nuclear
spin-lattice relaxation rate.\cite{matano08,parish08} However,
since most experiments are only sensitive to the magnitude of the gap of superconducting (SC) order, a direct {\it
phase}-sensitive experiment is essential to map out the complete
picture of the pairing symmetry.  So far there is no proposed direct
phase-sensitive experiment for iron pnictides similar to the dc
superconducting quantum interference device (SQUID) interferometer for the cuprates. The difficulty arises from the non-trivial phase structure of the order parameter in $k$ space. One possible phase-sensitive experiment is Andreev spectroscopy in the normal metal to superconductor (NS) junction. Unfortunately, two recent experiments~\cite{chen08,shan08} give seemingly conflicting results and detailed theoretical study shows indistinguishable features between an usual $s$-wave and sign-changed $s$-wave symmetries.\cite{linder08}

%In fact, the tunneling spectroscopy of normal metal-superconductor (NS) junctions is just one of such phase-sensitive experiments for the pairing symmetry. For instance, a sharp zero-bias conductance peak (ZBCP) due to the formation of zero-energy Andreev surface bound states would show up when the superconducting order parameter contains a node with its sign changed across the interface along the tunneling direction~\cite{hu94}. Unfortunately, the results in two recent NS-junction experiments~\cite{chen08,shan08} on iron pnictides seem to be conflicting with each other and hence it is still elusive whether the pairing symmetry is nodal or nodeless. In addition, although recent detailed theoretical study on the conductance spectra of NS junction involving iron pnictides shows different features for different pairing states, no sharp distinct feature is seen especially between conventional $s$-wave and $s_{x^2 y^2}$-wave symmetries.

\begin{figure}[tbh]
\begin{center}
\includegraphics[width=5.4cm]{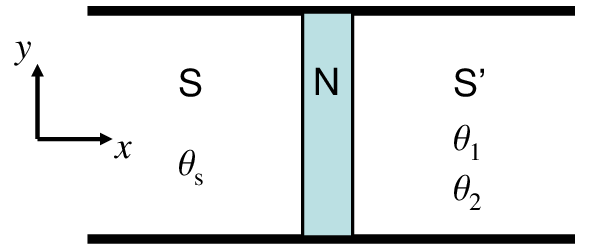}
\caption{(Color online) A schematic plot of the $\text{SNS}^\prime$ junction.}
\label{fig:SNSjunc}
\end{center}
\end{figure}

In this paper, we theoretically consider two types of Josephson
junctions which have novel properties uniquely
associated with a sign-changed $s$-wave SC order as opposed to
other (singlet) pairing symmetries.
Specifically, the first type of
junction we consider is a trilayer SC device where the iron-based
superconductor is sandwiched by two other layered, $s$-wave
superconductors [see inset of Fig.~\ref{fig:piJunc} (b)]. With
certain chosen FSs of the two outside layers which couple
stronger, respectively, to the hole and electron pockets of the iron
pnictide, due to momentum conservation, it is shown that
sign-changed $s$-wave pairing symmetry uniquely gives rise to a
$\pi$-junction behavior.\cite{berg08,harlingen95}
The second type is a single-band
superconductor-normal metal (or insulator)-iron pnictide
($\text{SNS}^\prime/\text{SIS}^\prime$) junction (see
Fig.~\ref{fig:SNSjunc}).
Based on the similar physics of Andreev reflection
at the interface between a normal metal and a superconductor,\cite{hu94,ghaemi08}   we demonstrate that, by adopting a minimal two-orbital model~\cite{seo08,raghu08} to include the multi-orbital effect and complex FSs,
the non-trivial phase structure of the sign-changed $s$-wave symmetry shows up
in the profile of the quasi-particle (QP) local density of states (LDOS) in the normal region of the junction (see Fig.~\ref{fig:SNSldos}): the sign-changed $s$-wave symmetry state supports in-gap bound state solutions.

%The scattering processes of Andreev reflection at the interface between a normal metal and a superconductor \cite{hu94,ghaemi08} may lead to the emergence of in-gap bound state solutions in the quasi-particle (QP) spectrum of a $\text{SNS}^\prime/\text{SIS}^\prime$ junction. By adopting a minimal two-orbital model~\cite{seo08,raghu08} to include the multi-orbital effect and complex FSs, the non-trivial phase structure of the sign-changed $s$-wave symmetry shows up in the profile of the QP local density of states (LDOS) in the normal region (see Fig.~\ref{fig:SNSldos}).

$Novel$ $\pi$-$junction$. We propose a composite Josephson
junction in which $\pi$-junction behavior can occur based on the
unusual phase structure of the $s_{x^2y^2}$-wave pairing.
A $\pi$-junction defines the situation when the Josephson coupling $J$ between two superconductors becomes {\it real} and {\it negative} (with no spontaneous or
explicit time reversal symmetry breaking). In other words, the ground
state energy (GSE) is minimized as the phase difference $\Phi$ between two
superconductors is ``$\pi$'', in contrast to the case of a ``0''-junction. The occurrence of the $\pi$-shift behavior can be usually due to magnetic ordering, strong correlation effects near the tunneling interface,\cite{berg08} or non-trivial phase structure of the SC order parameter such as $d_{x^2-y^2}$ pairing symmetry.\cite{harlingen95}

Unlike these common designs, our proposed
%As shown in the inset of Fig.~\ref{fig:piJunc} (b), the
junction [see the inset of Fig.~\ref{fig:piJunc} (b)] is composed of  an iron-pnictide ($\text{S}_m$) sandwiched by a top and a bottom quasi-2D $s$-wave superconductors ($\text{S}_t$ and $\text{S}_b$). The interface between any two
superconductors is an insulating thin film playing the role of a tunneling barrier. The key requirement for the top and bottom superconducting materials is that {\it the Cooper-pair tunneling probability is stronger into the hole (electron) pockets for the top (bottom) or vice versa.} This could be engineered to be due to
the normal state FSs of the top and bottom superconductors. One possible way to achieve this condition is to select a small FS and a large FS around $\Gamma$ point for the top layer and the bottom layer, respectively [see
Fig.~\ref{fig:piJunc} (a)], provided the in-plane (perpendicular to the tunneling direction) momentum is conserved ideally after tunneling.

\begin{figure}[thb]
\begin{center}
\includegraphics[scale=0.75]{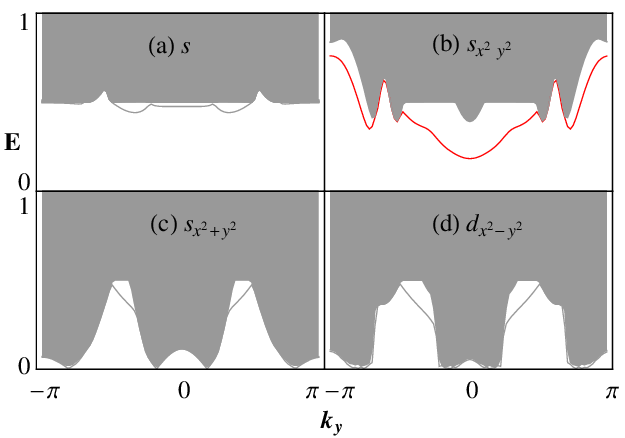}
\caption{(Color online) QP spectrum in the $\text{SNS}^\prime$
junction involving the iron-pnictide with various pairing
symmetries. The red line represents Andreev bound states in the case
of $s_{x^2 y^2}$-wave pairing. The used parameters are $U=0$,
$\tilde{\mu}=-2$, $\Delta_s=0.5$, $g_{S}=g_{S^\prime}=1$, $t_1=-1$,
$t_2=1.3$, $t_3=-0.85$, $t_4=-0.85$, $\mu=1.58$, $n_m=1$, and the
pairing strength is, respectively, (a) $\Delta_0=0.5$ (n=201), (b)
$\Delta_{s2}=0.5$ (n=201), (c)$\Delta_{s1}=0.5$ (n=201), and (d)
$\Delta_d=0.5$ (n=601) in energy units, $t_s\equiv 1$.}
\label{fig:SNSspec}
\end{center}
\end{figure}

\begin{figure}[thb]
\begin{center}
\includegraphics[scale=0.75]{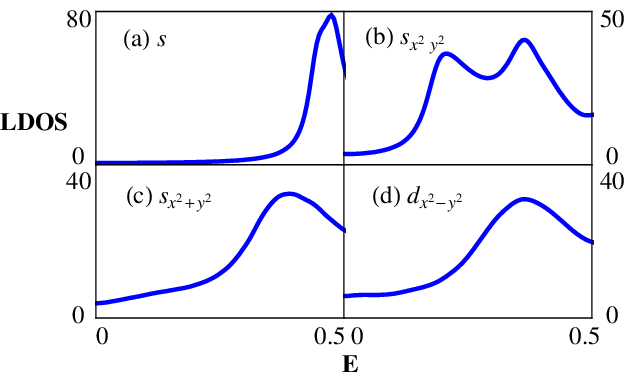}
\caption{(Color online)
QP-LDOS at $x=0$ in the $\text{SNS}^\prime$ junction involving
iron-pnictide with various pairing symmetries.  The used parameters
are the same as in drawing Fig.~\ref{fig:SNSspec}, except that $n=121$ for (a)-(c)
and $n=161$ for (d).}
\label{fig:SNSldos}
\end{center}
\end{figure}

A simple mean-field model Hamiltonian for this
trilayer junction can be of the form,  
$H_{J}=H_{t}+H_{m}+H_{b}+H_{T}$,
where $H_{\eta}=\sum_{\bk}\hat{\psi}^\dagger_{\eta}[(\varepsilon_{\eta,\bk}-\mu)\sigma_3+
\Delta_{\eta}e^{i\theta_{\eta}}\sigma_+
+\Delta_{\eta}e^{-i\theta_{\eta}}\sigma_-]\hat{\psi}_{\eta}$ for
$\eta=t,b$. $\varepsilon_{\eta,\bk}$ has the form of $-2t_{\eta}(\cos
k_x+\cos k_y)+\epsilon_{\eta}$, $\sigma_{\pm}=(\sigma_1\pm i\sigma_2)/2$, and $\hat{\psi}_{\eta}$ is the usual
Nambu spinor,
$\hat{\psi}^\dagger_{\eta}=(c^\dagger_{\eta,\bk,\uparrow},c_{\eta,
  -\bk,\downarrow})$. The difference between the top
and bottom SC phases is gauge invariant for the whole junction and is set to be
$\Phi=\theta_t-\theta_b$. $H_m$, the Hamiltonian of the iron pnictide,
is shown
%the same as $H_{S^\prime}
in Eq.~(\ref{eq:hsp}) transformed into
momentum space with band parameters and non-vanishing $\Delta_{s2}=\Delta e^{i\theta_m}$ given in the caption of Fig.~\ref{fig:SNSspec}. The tunneling
Hamiltonian, $H_T$, which connects neighboring layers, takes the
simple form:
$H_T=\sum_{\mathbf{p},\bk,\eta}\hat{\psi}^\dagger_{\eta}(\mathbf{p})\hat{h}_{T,\eta}\hat{\psi}_m(\bk)$, where
$\hat{h}_{T,\eta}$ is a $2\times 4$ matrix, \be
\left(\begin{array}{cccc}
    g_\eta & 0 & g_\eta & 0 \\
    0 & -g_\eta & 0 & -g_\eta
\end{array}
\right)\delta_{\mathbf{p},\bk}
\ee
and the spinor $\hat{\psi}^\dagger_m=(c^\dagger_{1,\bk,\uparrow},c_{1,-\bk,\downarrow},
c^\dagger_{2,\bk,\uparrow},c_{2,-\bk,\downarrow})$ describes the iron pnictide.
Note that we have assumed that the dispersion along $z$-axis is irrelevant and negligible in quasi-2D materials.

For demonstration purpose, we choose parameters
$t_{t}=t_{b}=1,\epsilon_{t}=4.78,\epsilon_{b}=1.88$,
$g_{t}=g_{b}=0.01$,
$\Delta_t=0.5,\Delta_b=0.4,\Delta=0.5$, and $\theta_m=0$. The stacked FSs
in the first Brillouin zone from each layer is shown in
Fig.~\ref{fig:piJunc} (a). Now, it is easy to diagonalize $H_J$ and
the ground-state energy for this mean-field Hamiltonian is simply the
sum of all QP eigen-energies below $E=0$. As presented in
Fig.~\ref{fig:piJunc} (b), the ground state energy per site relative
to the energy of $\Phi=0$ decreases as a function of $\Phi$ with its
minimum located at ``$\pi$''. The physics of this result can be easily captured by the perturbation calculations of GSE, which give, up to second order of $g^2_{\eta}$, $-J_{tm}\cos(\theta_t-\theta_m)-J_{mb}\cos(\theta_m-\theta_b)$,
where the Josephson couplings,$J_{tm}>0,J_{mb}<0$ (the sign difference between the Josephson couplings is due to $\pi$  phase difference between electron-like and hole-like FSs), have similar form of the textbook derivation.\cite{deGennes} It is obvious to see that the global minimum reaches at
$|\theta_t-\theta_b|=\pi$. The overall junction shows ``$\pi$'' behavior.

%This $\pi$-shift behavior is unchanged provided $0\le\theta_m<\pi/2$, which can %be understood as follows...
%This numerical result is indeed
%consistent with the perturbative calculations, where in the limit of
%$|g_{\alpha}/\Delta_{\alpha}|\ll 1$, $\Delta
%E_{GS}/N\sim-J_{eff}\cos(\Phi)$, with a ``negative'' Josephson
%coupling density $J_{eff}\sim -c(g_t g_b)^2/\Delta_t\Delta_b$
%[$c\sim\Cal{O}(1)$] across the middle layer and tunneling barriers.

Some comments on the experimental realization are in order. First,
``large'' or ``small'' FS is meaningful only when the lattice constant
$a$ is equal or comparable to the iron pnictides, where the
nearest-neighbor Fe-Fe distance is around 2.85\AA. Second, to make the
tunneling processes reasonably dominated by in-plane momentum
conservation, quasi-2D, $s$-wave SC materials should be used for the
top and bottom layers due to their irrelevant dispersion along $z$
direction and the epitaxial growing technique may be useful for making the coherent-tunnel interfaces. Based on these considerations, some plausible candidates
for the large FS are, for instance, $\text{MgB}_2$ from its $\pi$ band
with $a\sim 3$\AA~or thin film of Beryllium with $a\sim 2.3$\AA; for
the small FS, it could be 2H-$\text{NbSe}_2$, where $a\sim 3.45$\AA.\cite{candidates}

$SNS^\prime/SIS^\prime$ $junctions$. A further feature of the $s_{x^2y^2}$-SC is revealed by considering a Josephson junction which connects, on one side, a single-band $s$-wave superconductor, through a normal metal to, on the other side, an iron-based superconductor.\cite{linder09}
%this junction device
We assume its QP spectrum is well approximated by a BCS type mean-field
Hamiltonian subject to an inhomogeneous pairing field along the tunneling  direction. Ignoring the $z$-axis for simplicity, the 2D model Hamiltonian of this junction reads,
\ba
\Cal{H} &=&
\Theta(-x-\frac{d}{2})H_S+\Theta(\frac{d}{2}-|x|)H_{N}+
\Theta(x-\frac{d}{2})H_{S^\prime} \nonumber \\
&+& H_{T}, \\
H_S &=& -t_S\sum_{\langle\br\brprime\rangle,\sigma}c_{\br,\sigma}^\dagger
c_{\brprime,\sigma}-\tilde{\mu}\sum_{\br,\sigma}c^\dagger_{\br,\sigma}
c_{\br,\sigma} \nonumber \\
&+& \sum_{\br}(\Delta_s c^\dagger_{\br,\uparrow} c^\dagger_{\br,\downarrow}+h.c.), \\
H_N &=& -t_N\sum_{\langle\br\brprime\rangle,\sigma}f_{\br,\sigma}^\dagger
f_{\brprime,\sigma}+(U-\tilde{\mu})\sum_{\br,\sigma}f^\dagger_{\br,\sigma}
f_{\br,\sigma},
\ea
where $\Theta(x)$ is the Heaviside function, the shifted chemical potential,  $\tilde{\mu}=\mu-\epsilon$, guarantees a partially filled band, and $U$ denotes barrier potential. To mimic the iron pnictide FS, we adopt a two-orbital exchange coupling model.~\cite{seo08} This leads to a somewhat complicated form of $H_{S^\prime} = H_0+H_\Delta$, where we separate it into the band structure and pairing field parts,
\ba
H_0 &=& \sum_{\br,\sigma}[-t_1 c^\dagger_{1,\br,\sigma}c_{1,\br+\hat{x},\sigma}
-t_2 c^\dagger_{1,\br,\sigma}c_{1,\br+\hat{y},\sigma}+h.c.] \nonumber \\
&+& \sum_{\br,\sigma}[-t_2 c^\dagger_{2,\br,\sigma}c_{2,\br+\hat{x},\sigma}
-t_1 c^\dagger_{2,\br,\sigma}c_{2,\br+\hat{y},\sigma}+h.c.]  \nonumber \\
&+& \sum_{\langle\langle\br\brprime\rangle\rangle}\sum_{\alpha,\sigma}-t_3 c^\dagger_{\alpha,\br,\sigma}c_{\alpha,\brprime,\sigma}-\mu\sum_{\alpha,\br,\sigma}c^\dagger_{\alpha,\br,\sigma}
c_{\alpha,\br,\sigma} \nonumber \\
&-&\sum_{\langle\langle\br\brprime\rangle\rangle}\sum_{\sigma}[t_4
e^{i\frac{\pi}{2}[(x^\prime-x)+(y^\prime-y)]} c^\dagger_{1,\br,\sigma}c_{2,\brprime,\sigma}
+h.c.], \nonumber \\
H_{\Delta}&=&  [\sum_{\alpha,\br}\Delta_0 c^\dagger_{\alpha,\br,\uparrow}
c^\dagger_{\alpha,\br,\downarrow}+ \sum_{\alpha,\langle\langle\br\brprime\rangle\rangle}
\frac{\Delta_{s2}}{4}c^\dagger_{\alpha,\br,\uparrow}
c^\dagger_{\alpha,\brprime,\downarrow} \nonumber \\
&+& \sum_{\alpha,\langle\br\brprime\rangle}\frac{\Delta_{d}e^{i\pi(\alpha-1)}}{4}
\phi_{\br\brprime}(c^\dagger_{\alpha,\br,\uparrow}
c^\dagger_{\alpha,\brprime,\downarrow}-c^\dagger_{\alpha,\br,\downarrow}
c^\dagger_{\alpha,\brprime,\uparrow})  \nonumber \\
&+& \sum_{\alpha,\langle\br\brprime\rangle}\frac{\Delta_{s1}}{4}
(c^\dagger_{\alpha,\br,\uparrow}
c^\dagger_{\alpha,\brprime,\downarrow}-c^\dagger_{\alpha,\br,\downarrow}
c^\dagger_{\alpha,\brprime,\uparrow})]+h.c.
\label{eq:hsp}
\ea
where $\alpha=1,2$ correspond to
$d_{xz}$ and $d_{yz}$ orbitals, respectively, and the hopping
parameters are given in Fig.~\ref{fig:SNSspec}.
$\langle\langle\br\brprime\rangle\rangle$ represents a next
nearest-neighbor pair and
%newly edited
$\phi_{\br\brprime}=1(-1)$ as $\br-\brprime=\pm\hat{x}(\hat{y})$.
%%%
All pairing fields considered in
$\Cal{H}$ are set to be real.\cite{realcond}
$\Delta_0,\Delta_{s1},\Delta_{s2}$, and $\Delta_{d}$ correspond to
intra-orbital on-site, $\cos k_x+\cos k_y$, $\cos k_x\cos k_y$, and
$\cos k_x-\cos k_y$ pairing strength, respectively, and inter-orbital
pairing is ignored due to its small contribution as discussed
in Ref.~\onlinecite{seo08}. Finally, $H_T$ describes the tunneling
amplitudes across two interfaces around $\pm d/2$. It can be written
as \be
H_T=g_{S}\sum_{\sigma}c^\dagger_{\br_L,\sigma}f_{\brprime_L,\sigma}
+g_{S^\prime}\sum_{\alpha,\sigma}c^\dagger_{\alpha,\br_R,\sigma}
f_{\brprime_R,\sigma}+h.c., \ee where $\br_L,\brprime_L$
($\br_R,\brprime_R$) are understood to be coordinates across the
left (right) interface.

Before proceeding to compute the QP spectrum and the corresponding
LDOS, it is important to realize that $\Cal{H}$ is
particle-hole symmetric under $c_{\br,\uparrow}\rightarrow
c^{\dagger}_{\br,\downarrow},c_{\br,\downarrow}\rightarrow
-c^\dagger_{\br,\uparrow}$ (for all fermion operators),
and hence its spectrum should be symmetric with respect to zero-energy. Taking advantage of the translational symmetry
transverse to the tunneling direction $x$, $\Cal{H}$ can be
further decomposed into a sum of 1D Hamiltonians by partially
Fourier-transforming along (100) surface ($y$ direction). As a
consequence, the whole system is mapped onto an 1D effective lattice
in the form of $\Cal{H}=\sum_{k_y}H_{1D}(k_y)$. Basically,
this transformation results in effective chemical potential and
pairing fields with $k_y$-dependence.

We diagonalize the model Hamiltonian $H_{1D}(k_y)$ for $-\pi\le
k_y<\pi$ on the Nambu basis $\Psi$, where in numerical calculations
we take the total number of the $1D$ lattice sites $n$ sufficiently
large with open boundary conditions; the normal metal is always set
in the middle of system and it is $n_m$-site wide. This length is always much less
than the SC coherence length. In
Fig.~\ref{fig:SNSspec}, we show numerical results of the zero
temperature QP spectrum as a function of $k_y$ for various pairing
symmetries of the iron-pnictides at $\mu=1.58$ (electron doped). In
addition, to visualize the Andreev bound states, in
Fig.~\ref{fig:SNSldos} we also compute each corresponding
QP-LDOS as a function of position and energy,
$D(x,\omega)=\sum_{k_y,i}|\Psi_i(x,k_y)|^2\delta(\omega-\epsilon_i)$,
where $i$ denotes the $i$th eigenfunction. The magnitude of the bulk
SC order parameters on both sides of the junction is taken to be the
same. The ratio of the gap to the half
band-width of the spectrum is around 0.02 and the coherence length
is estimated as $\xi\sim v_F/\Delta_s\sim 4a$.

As clearly seen in Fig.~\ref{fig:SNSspec}, the presence of the in-gap
Andreev bound states with significant weight in different $k_y$
channels for the extended $s$-wave $\cos k_x\cos k_y$ ($s_{x^2 y^2}$) pairing
symmetry is a sharp feature distinguishing it
from other pairing symmetries. Although in addition to (gray-color filled) continuum states there are discrete energy levels
for $s$, $s_{x^2+y^2}$, and $d_{x^2-y^2}$ pairing symmetries,
they are either near the maximum gap edge ($s$) or only appear in certain range of $k_y$ ( for $s_{x^2+y^2}$, $d_{x^2-y^2}$). Especially for the latter case,
due to the presence of nodal points on the electron or the
hole pockets, the contribution from scattering-state
can easily overwhelm that from the bound states and can lead to qualitatively different QP-LDOS from the case of
$s_{x^2 y^2}$-wave, where a sharp peak appears at the positive
subgap energy.

Furthermore, two observations deserve mentioning. First, the features shown in Figs.~\ref{fig:SNSspec} and
\ref{fig:SNSldos} do not change much for different doping levels
as long as the doping concentration is not large enough so that the
Fermi surfaces pass the nodal line of $\cos k_x\cos k_y$, i.e.,
$k_x=\pm\pi/2$ and $k_y=\pm\pi/2$. Second, if the barrier potential
$U$ is greater than the difference between $\mu$ and the band
bottom, the normal region becomes insulating. This
moves the subgap peak in the $s_{x^2 y^2}$ LDOS closer to the gap edge without
destroying it.

Can we understand the presence of such non-trivial bound states for
the $s_{x^2 y^2}$-wave pairing in a simple way? A physical
insight for this junction involving such an unconventional symmetry
can be obtained by treating the {\it bands} at the
electron and hole pockets in the iron-pnictide as independent of each
other.\cite{ng08} Consequently, a simple description for the junction based on
Bogoliubov-de Gennes (BdG) equations reads:
\begin{equation}[\hat{H}_{0,\lambda}\sigma_3+\Delta_\lambda(\br)\sigma_{+}+
\Delta_\lambda^*(\br)\sigma_{-}]\psi(\br)=
\epsilon_{\lambda}\psi(\br),\end{equation}where $\lambda=1,2$ are the {\it band}
indices, and $\sigma_{i}(i=1,2,3)$ are the Pauli matrices with
$\sigma_{\pm}=(\sigma_1\pm i\sigma_2)/2$ acting in the Nambu space,
$\psi=[\tilde{u}_{\lambda}(\br),\tilde{v}_{\lambda}(\br)]^t$.
$\hat{H}_{0,\lambda}$ contains the band information of non-interacting electrons and $\Delta_\lambda(\br)$ corresponds to the pairing field
in the same band $\lambda$. Along the tunneling direction $x$, the
inhomogeneous $\Delta_{\lambda}(\br)$ is modeled by
$\Delta_\lambda(x)=\Delta_s e^{i\theta_{s}}$ as $x<0$ and
$\Delta_\lambda(x)=\Delta_\lambda e^{i\theta_\lambda}$ as $x>0$.
For simplicity, we set $\theta_{s}=0$ hereafter and keep in mind
that for the sign-changed $s$-wave symmetry,
$\theta_1=\theta_2+\pi$. For electrons near FSs, it is valid to
linearize BdG equations within WKJB approximation, $\psi\sim
e^{i\bk_F\cdot\br}\phi(\br)$, $\phi(\br)=(u(\br),v(\br))^t$ and then
the BdG equations are now reduced to the form of 1D Dirac equation
if we further take the advantage of translational symmetry in
transverse direction, \be \label{eq:1Ddirac}
[-iv_{Fx}\partial_x\sigma_3+\Delta_\lambda(x)\sigma_{+}+
\Delta_\lambda^*(x)\sigma_{-}]\phi(x)= \epsilon_{\lambda}\phi(x).
\ee After straightforward calculations with trial bound state
solutions, $u (v)\sim u_0 (v_0)e^{-\gamma_{\pm}x}$, as studied
in Ref.~\onlinecite{ng08}, the discrete energy
level within the gap is given by
$E_0=\pm(\Delta_s\Delta_\lambda\sin\theta_\lambda)/
\sqrt{\Delta_s^2+\Delta_\lambda^2-2\cos\theta_\lambda\Delta_s\Delta_\lambda}$,
provided $\cos\theta_\lambda<$
min($\Delta_\lambda/\Delta_s,\Delta_s/\Delta_\lambda$). The pair of
solutions with eigenvalues symmetric with respect to zero-energy
follows from the particle-hole symmetry of the BdG equations. It is
clear to see that when superconductors on both sides of the junction
are in phase ($\theta_\lambda=0$) no bound state solution is found,
while when they are out of phase ($\theta_\lambda=\pi$) there are
doubly-degenerate zero modes.\cite{ng08}

The significance of this simple result is that as long as $\theta_1=0$
(or $\theta_2=0$), there are always zero modes trapped in the normal
region of the junction involving iron pnictides with sign-changed
$s$-wave pairing symmetry. However, in a more realistic system with
band structure such as our $\text{SNS}^\prime$ junction it usually
introduces finite effective mass for the band electrons, which
destroys the validity of using linearized Eq.~(\ref{eq:1Ddirac})
(where the effective mass goes to infinity). As a result, the
to-be-degenerate zero modes split~\cite{tsai08} as we see in
Fig.~\ref{fig:SNSspec} (only $E>0$ shown). The same argument is also applicable when we
further consider the effect brought by $k_y$ in our $H_{1D}(k_y)$.

%$Summary$--We have studied Andreev bound states in
%$\text{SNS}^\prime$ ($\text{SIS}^\prime$) junction involving the
%iron pnictides. The non-trivial phase structure of $s$-wave $\cos
%k_x\cos k_y$ SC order leads to the formation of the in-gap bound
%states, and hence exhibits a sharp sub-gap peak in the QP-LDOS,
%which can be taken as a unique feature to distinguish from other
%possible pairing symmetries in the iron pnictides. In addition, the
%proposed trilayer Josephson junction offers the unconventional
%sign-changed $s$-wave pairing a new experimental detectable
%signature, namely,  a novel $\pi$-junction.

$Acknowledgements$. We thank E.~Berg and  C.~Fang for stimulating
and useful discussions. JPH and WFT are supported by NSF 
Grant No. PHY-0603759 and DXY is supported by NSF Grand No. DMR-0804748.

%\vspace{-0.4cm}


\begin{thebibliography}{99}
\bibitem{realcond} We neglect the possibility of a spontaneously time reversal symmetry broken ground state.
\end{thebibliography}
\end{document}